\RequirePackage{fix-cm}
\documentclass[twocolumn]{svjour3}          
\smartqed  
\usepackage{graphicx}
\usepackage[pdfborder={0 0 0},colorlinks=true]{hyperref}

\def\beq{\begin{equation}}
\def\eeq{\end{equation}}
\def\bea{\begin{eqnarray}}
\def\eea{\end{eqnarray}}

\def\ra{\rangle}
\def\la{\langle}

\def\k{\mathbf{k}}

\newcommand{\sgn}[1] {\mathrm{sgn}\left({#1}\right)}

\newcommand{\ii}{{\mathrm{i}}}

\newcommand{\nn}{\nonumber}

\newcommand{\Tmat}{\mathcal{T}}
\newcommand{\Uimp}{\mathcal{U}}

\newcommand{\iM}{\mathcal{I}}
\newcommand{\jM}{\mathcal{J}}
\newcommand{\imp}{\mathrm{imp}}

\newcommand{\GammaN}{\Gamma}
\newcommand{\sigmaN}{\sigma}
\newcommand{\etaN}{\eta}

\begin{document}

\title{Uniform impurity scattering in two-band $s_\pm$ and $s_{++}$ superconductors}

\author{M.M. Korshunov \and Yu.N. Togushova \and O.V. Dolgov}


\institute{M.M. Korshunov \at
              \email{mkor@iph.krasn.ru} \\
              L.V. Kirensky Institute of Physics, Krasnoyarsk 660036, Russia
           \and
           M.M. Korshunov and Yu.N. Togushova \at
              Siberian Federal University, Svobodny Prospect 79, Krasnoyarsk 660041, Russia
           \and
           O.V. Dolgov \at
              Max-Planck-Institut f\"{u}r Festk\"{o}rperforschung, D-70569 Stuttgart, Germany\\
              Lebedev Physical Institute RAN, Moscow, Russia
}

\date{Received: date / Accepted: date}

\maketitle

\begin{abstract}
The $s_\pm$ and $s_{++}$ models for the superconducting state are subject of intense studies regarding Fe-based superconductors. Depending on the parameters, disorder may leave intact or suppress $T_c$ in these models. Here we study the special case of disorder with equal values of intra- and interband impurity potentials in the two-band $s_\pm$ and $s_{++}$ models. We show that this case can be considered as an  isolated point and $T_c$ there 
has maximal damping
for a wide range of parameters.
\end{abstract}

\section{Introduction}

Fe-based materials - pnictides and chalcogenides - represent a new class of unconventional superconductors with high transition temperatures~\cite{reviews}. While the mechanism of superconductivity is still a mystery, the main candidates are spin or orbital fluctuations. Except for the extreme hole and electron dopings, the Fermi surface consists of two or three hole pockets around the $\Gamma=(0,0)$ point and two electron pockets around the $M=(\pi,\pi)$ point in the 2-Fe Brillouin zone. Scattering between them with the large wave vector results in the enhanced antiferromagnetic fluctuations, which promote the $s_\pm$ type of the superconducting order parameter that change sign between electron and hole pockets~\cite{reviews}. On the other hand, bands near the Fermi level have mixed orbital content and orbital fluctuations enhanced by the electron-phonon interaction may lead to the sign-preserving $s_{++}$ state~\cite{kontani,bang}. However, most experimental data including observation of a spin-resonance peak in inelastic neutron scattering, the quasiparticle interference in tunneling experiments, and NMR spin-lattice relaxation rate are in favor of the $s_\pm$ scenario~\cite{reviews}.

The $s_\pm$ and $s_{++}$ states are expected to behave differently subject to the disorder~\cite{AG,golubov97,Ummarino2007,kontaniimp,kontaniimspm,efremov,Efremov2013,Dolgov2013,Korshunov2014}.
In general, $s_{++}$ ($s_\pm$) state should be stable (fragile) against a scattering on a nonmagnetic impurities~\cite{AG,golubov97,Ummarino2007}. Detailed studies revealed that $T_c$ stays finite in the presence of nonmagnetic disorder in the following cases: i) $s_{++}$ state~\cite{kontaniimp,kontaniimspm}, ii) $s_\pm \to s_{++}$ transition for the sizeable intraband attraction in the two-band $s_\pm$ model in the strong-coupling $\Tmat$-matrix approximation~\cite{efremov} and via the numerical solutions of the Bogoliubov-de Gennes equations~\cite{Yao2012,Chen2013}, iii) an unitary limit~\cite{dolgovkulic}. Magnetic impurities leave $T_c$ finite~\cite{Korshunov2014} in the case of 1) $s_\pm$ superconductor with the purely interband impurity scattering, 2) $s_{++}$ state with the purely interband scattering due to the $s_{++} \to s_\pm$ transition, and 3) the unitary limit for both $s_{++}$ and $s_\pm$ states independent on the exact form of the impurity potential. But even if $T_c$ is suppressed, its behavior may differ from the Abrikosov-Gor'kov (AG) theory for the single-band superconductors~\cite{AG}, which states that $T_c$ is determined by the expression $\ln T_{c0}/T_{c} = \Psi(1/2+ \Gamma/2\pi T_c) - \Psi(1/2)$, where $\Psi(x)$ is the digamma function, $\Gamma$ is the impurity scattering rate, and $T_{c0}$ is the critical temperature in the absence of impurities~\cite{AG}.

The choice of the ``proper'' theory for disorder effects in iron-based materials is severely complicated by the fact that the exact form of the impurity potential is not known. In such a situation it is instructive to theoretically explore as many situations as possible. Here we focus on a special case of a uniform impurity potential
with equal intra- and interband components. We consider two-band models for the isotropic $s_\pm$ and $s_{++}$ superconductors with either nonmagnetic or magnetic impurities within the self-consistent $\Tmat$-matrix approximation following approach from Refs.~\cite{efremov,Korshunov2014}.

\section{General equations and their analysis}

We employ the Eliashberg approach for multiband superconductors~\cite{allen} and calculate the $\xi$-integrated Green's functions $\hat{\mathbf{g}}(\omega_n) = \int d \xi \hat{\mathbf{G}}(\k, \omega_n) =
\left(
\begin{array}{cc}
\hat{g}_{an} & 0 \\
0 & \hat{g}_{bn}
\end{array}
\right)$,
%
%
%
%
where $\xi_{\alpha, \k} = \mathbf{v}_{\alpha, F} (\k-\k_{\alpha, F})$ is the linearized dispersion, $\k_{\alpha, F}$ is the Fermi momentum, $\hat{g}_{\alpha n} = g_{0\alpha n} \hat{\tau}_{0}\otimes \hat{\sigma}_{0} + g_{2\alpha n} \hat{\tau}_{2}\otimes \hat{\sigma}_{2}$, indices $a$ and $b$ correspond to two distinct bands, index $\alpha = a,b$ denote the band space, Pauli matrices define Nambu ($\hat{\tau}_{i}$) and spin ($\hat{\sigma}_{i}$) spaces, $\hat{\mathbf{G}}(\k,\omega_n) = \left[\hat{\mathbf{G}}_0^{-1}(\k,\omega_n) - \hat{\mathbf{\Sigma}}(\omega_n)\right]^{-1}$ is the matrix Green's function for a quasiparticle with momentum $\k$ and the Matsubara frequency $\omega_n = (2 n + 1) \pi T$ defined in the band space and in the combined Nambu and spin spaces, $\hat{G}_0^{\alpha \beta}({\mathbf{k}},\omega_n)=[ \ii \omega_n \hat{\tau}_{0}\otimes \hat{\sigma}_{0}-\xi_{\alpha \k}\hat{\tau}_{3}\otimes \hat{\sigma}_{0}]^{-1} \delta_{\alpha \beta}$ is the bare Green's function,
$\hat{\mathbf{\Sigma}}(\omega _{n}) = \sum_{i=0}^{3} \Sigma_{\alpha \beta}^{(i)}(\omega_n)\hat{\tau}_i$ is the self-energy matrix, $g_{0\alpha n}$ and $g_{2\alpha n}$ are the normal and anomalous $\xi$-integrated Nambu Green's functions,
\beq
g_{0\alpha n}=-\frac{\ii \pi N_{\alpha} \tilde{\omega}_{\alpha n}}{\sqrt{\tilde{\omega}_{\alpha n}^{2}+\tilde{\phi}_{\alpha n}^{2}}}, \;\;\; g_{2\alpha n}=-\frac{\pi N_{\alpha} \tilde{\phi}_{\alpha n}}{\sqrt{\tilde{\omega}_{\alpha n}^{2}+\tilde{\phi}_{\alpha n}^{2}}},
\label{g}
\eeq
depending on the density of states per spin of the corresponding band at the Fermi level $N_{a,b}$ and on renormalized (by the self-energy) order parameter $\tilde{\phi}_{\alpha n}$ and frequency $\tilde{\omega}_{\alpha n}$,
\begin{eqnarray}
\ii \tilde\omega_{a n} &=& \ii \omega_n -  \Sigma_{0a}(\omega_n) - \Sigma_{0a}^{\imp}(\omega_n), \label{eq.omega.tilde} \\
\tilde\phi_{a n} &=& \Sigma_{2a}(\omega_n) + \Sigma_{2a}^{\imp}(\omega_n). \label{eq.phi.tilde}
\end{eqnarray}
It is also convenient to introduce the renormalization factor $Z_{\alpha n} = \tilde{\omega}_{\alpha n} / \omega_n$ that enters the gap function $\Delta_{\alpha n} = \tilde{\phi}_{\alpha n} / Z_{\alpha n}$. The self-energy due to the spin fluctuation interaction is then given by
\bea
\Sigma_{0\alpha}(\omega_n) &=& T \sum\limits_{\omega_n',\beta} \lambda^{z}_{\alpha\beta} (n-n') \frac{g_{0\beta n}}{N_\beta}, \label{eq.DeltaN2} \\
\Sigma_{2\alpha}(\omega_n) &=& -T \sum\limits_{\omega_n',\beta} \lambda^{\phi}_{\alpha\beta}(n-n') \frac{g_{2\beta n}}{N_\beta},
\label{eq.DeltaN1}
\eea
The coupling functions $\lambda^{\phi,z}_{\alpha\beta}(n-n') = 2 \lambda^{\phi,z}_{\alpha\beta} \int\limits^{\infty}_{0} \frac{d\Omega \Omega B(\Omega)}{(\omega_n-\omega_{n'})^{2} + \Omega^{2}}$ depend on the normalized bosonic spectral function
$B(\Omega)$ used in Refs.~\cite{efremov,Efremov2013}.
While the matrix elements $\lambda^\phi_{\alpha \beta}$  can be positive (attractive) as well as negative (repulsive) due to the interplay between spin fluctuations and electron-phonon coupling~\cite{BS,parker}, the matrix elements $\lambda^z_{\alpha \beta}$ are always positive. For simplicity we set $\lambda^z_{\alpha \beta}=|\lambda^\phi_{\alpha \beta}|\equiv|\lambda_{\alpha \beta}|$ and neglect possible $\k$-space anisotropy in each order parameter $\tilde\phi_{\alpha n}$.

We use the $\Tmat$-matrix approximation to calculate the average impurity
self-energy $\hat{\mathbf{\Sigma}}^{\imp}$:
\begin{equation}
\hat{\mathbf{\Sigma}}^{\imp}(\omega_n) = n_{\imp} \hat{\mathbf{U}} + \hat{\mathbf{U}} \hat{\mathbf{g}}(\omega_n) \hat{\mathbf{\Sigma}}^{\imp}(\omega_n),
\label{eq.tmatrix}
\end{equation}
where $n_{\imp}$ is the impurity concentration.

\subsection{Nonmagnetic impurities}

First, we consider nonmagnetic disorder. Impurity potential matrix entering equation~(\ref{eq.tmatrix}) is defined as $\hat{\mathbf{U}} = \mathbf{U} \otimes \hat\tau_3$, where $(\mathbf{U})_{\alpha \beta} = \Uimp_{\mathbf{R}_{i}}^{\alpha \beta}$ with $\mathbf{R}_{i} = 0$ is the impurity site. For simplicity, we set intra- and interband parts of the potential equal to $v$ and $u$, respectively, so that $(\mathbf{U})_{\alpha \beta} = (v-u) \delta_{\alpha \beta} + u$. Relation between the two will be controlled by the parameter $\etaN$: $v = u \etaN$.

Apart from the general case, later we are going to examine the two important limiting cases: Born limit (weak scattering) with $\pi u N_{a,b} \ll 1$ and the opposite case of a very strong impurity scattering (unitary limit) with $\pi u N_{a,b} \gg 1$.

It is useful to introduce the generalized scattering cross-section
\beq
\sigmaN = \frac{\pi^2 N_a N_b u^2}{1 + \pi^2 N_a N_b u^2} \to \left\{
\begin{array}{l}
0, \mathrm{Born} \\
1, \mathrm{unitary}
\end{array}
\right.
\eeq
and the impurity scattering rate
\beq
\GammaN_{a,b} = \frac{2 n_{\imp} \sigmaN}{\pi N_{a,b}} \to \left\{
\begin{array}{l}
2 n_{\imp}\pi N_{b,a} u^2, \mathrm{Born} \\
2 n_{\imp}/\left( \pi N_{a,b} \right), \mathrm{unitary}
\end{array}
\right.
\eeq
Then equations on frequency~(\ref{eq.omega.tilde}) and order parameter~(\ref{eq.phi.tilde}) become
\begin{eqnarray}
\label{eq.omega.nonmagn}
\tilde{\omega}_{an} &=& \omega_n + \ii \Sigma_{0a}(\omega_n) \\
&+& \frac{\GammaN_a}{2 D} \left[ \sigmaN \frac{\tilde{\omega}_{an}}{Q_{an}} (1 - \etaN^2)^2 + (1 - \sigmaN) \left( \frac{N_a \tilde{\omega}_{an}}{N_b Q_{an}} \etaN^2 + \frac{\tilde{\omega}_{bn}}{Q_{bn}} \right) \right], \nn
\\
\label{eq.phi.nonmagn}
\tilde{\phi}_{an} &=& \Sigma_{2a}(\omega_n) \\
&+& \frac{\GammaN_a}{2 D} \left[ \sigmaN \frac{\tilde{\phi}_{an}}{Q_{an}} (1 - \etaN^2)^2 + (1-\sigmaN) \left( \frac{N_a \tilde{\phi}_{an}}{N_b Q_{an}} \etaN^2 + \frac{\tilde{\phi}_{bn}}{Q_{bn}} \right) \right]. \nn
\end{eqnarray}
where $Q_{\alpha n} = \sqrt{\tilde{\omega}_{\alpha n}^2 + \tilde{\phi}_{\alpha n}^2}$, $D = (1-\sigmaN)^2 + \sigmaN (1-\sigmaN) \left( 2 \frac{\tilde{\omega}_{an} \tilde{\omega}_{bn} + \tilde{\phi}_{an} \tilde{\phi}_{bn}}{Q_{an} Q_{bn}} + \frac{N_a^2 + N_b^2}{N_a N_b} \etaN^2 \right) + \sigmaN^2 (1 - \etaN^2)^2$.

Let's consider the main limits. Since in the Born approximation $\sigmaN \to 0$, then $D = 1$, $\GammaN_a = 2 n_{\imp} \pi N_b u^2$ and
\begin{eqnarray}
\tilde{\omega}_{an} &=& \omega_{n} + \ii \Sigma_{0a}(\omega_n) + \frac{\gamma_{aa}}{2} \frac{\tilde{\omega}_{an}}{Q_{an}} + \frac{\gamma_{ab}}{2} \frac{\tilde{\omega}_{bn}}{Q_{bn}},
\label{eq.omega.interBorn}
\\
\tilde{\phi}_{an} &=& \Sigma_{2a}(\omega_n) + \frac{\gamma_{aa}}{2} \frac{\tilde{\phi}_{an}}{Q_{an}} + \frac{\gamma_{ab}}{2} \frac{\tilde{\phi}_{bn}}{Q_{bn}},
\label{eq.phi.interBorn}
\end{eqnarray}
where $\gamma_{aa} = 2 \pi n_{\imp} N_a u^2 \etaN^2$ and $\gamma_{ab} = 2 \pi n_{\imp} N_b u^2$. Obviously, for the finite interband scattering $\gamma_{ab}$ (i.e. finite $\etaN$) different bands are mixed in equations. This leads to the AG-like suppression of $T_c$.

In the unitary limit $\sigmaN \to 1$, $\GammaN_a = 2 n_{\imp} / (\pi N_a)$, and we have to consider two cases.

I). Uniform impurity potential with $\etaN = 1$:
\begin{eqnarray}
\tilde{\omega}_{an} &=& \omega_n + \ii \Sigma_{0a}(\omega_n) + \frac{n_{\imp}}{\pi N_a N_b D_{uni}} \left[ N_a \frac{\tilde{\omega}_{an}}{Q_{an}} + N_b \frac{\tilde{\omega}_{bn}}{Q_{bn}} \right], \nn \label{eq.omega.uni.eta1.nonmag} \\
\tilde{\phi}_{a n} &=& \Sigma_{2a}(\omega_n) + \frac{n_{\imp}}{\pi N_a N_b D_{uni}} \left[ N_a \frac{\tilde{\phi}_{an}}{Q_{an}} + N_b \frac{\tilde{\phi}_{bn}}{Q_{bn}} \right], \nn
\label{eq.phi.uni.eta1.nonmag}
\end{eqnarray}
where $D_{uni} = 2 \frac{\tilde{\omega}_{an} \tilde{\omega}_{bn} + \tilde{\phi}_{an} \tilde{\phi}_{bn}}{Q_{an} Q_{bn}} + \frac{N_a^2 + N_b^2}{N_a N_b}$. Again, different bands are mixed so we have a suppression of $T_c$.

II). All other cases with $\etaN \neq 1$:
\begin{eqnarray}
\tilde{\omega}_{an} &=& \omega_n + \ii \Sigma_{0a}(\omega_n) + \frac{n_{\imp}}{\pi N_a} \frac{\tilde{\omega}_{an}}{Q_{an}}, \label{eq.omega.uni.nonmag} \\
\tilde{\phi}_{a n} &=& \Sigma_{2a}(\omega_n) + \frac{n_{\imp}}{\pi N_a} \frac{\tilde{\phi}_{an}}{Q_{an}}. \label{eq.phi.uni.nonmag}
\end{eqnarray}
We get the same result, as for the intraband impurities since the other band ($b$) does not contribute to the equations. Surprisingly, but here the Anderson theorem works independent of the gap signs in different bands. Thus, $T_c$ should be finite for arbitrary impurity concentration.

Here we conclude, that there is a special case of $T_c$ suppression in the unitary limit for the uniform impurity potential $\etaN = 1$. Such situation arise due to the structure of the denominator $D$ in equations~(\ref{eq.omega.nonmagn})-(\ref{eq.phi.nonmagn}). It vanishes for $\etaN = \sigmaN = 1$ and one has to accurately take the limit $\etaN \to 1$ first and only then put $\sigmaN \to 1$.
It is the $\etaN = 1$ case, that was considered in Ref.~\cite{bang}. For all other values of $\etaN$ (even for a slight difference between intra- and interband potentials) impurities are not going to affect the critical temperature.

\subsection{Magnetic impurities}

Now we switch to the magnetic disorder. Impurity potential for the non-correlated impurities can be written as $\hat{\mathbf{U}}=\mathbf{V} \otimes \hat{S}$, where
$ \hat{S} = \mathrm{diag}\left[\vec{\hat{\sigma}} \cdot \vec{S}, -(\vec{\hat{\sigma}} \cdot \vec{S})^{T}\right]$ is the $4 \times 4$ matrix with $(...)^{T}$ being the matrix transpose and $\vec{S} = \left( S_x, S_y, S_z \right)$ being the spin vector~\cite{ambeg}. The vector $\vec{\hat{\sigma}}$ is composed of $\tau$ matrices, $\vec{\hat{\sigma}} = \left( \hat{\tau}_1, \hat{\tau}_2, \hat{\tau}_3 \right)$. The potential strength is determined by $(\mathbf{V})_{\alpha \beta} = V_{\mathbf{R}_i = 0}^{\alpha \beta}$. For simplicity, intraband and interband parts of the potential are set equal to $\iM$ and $\jM$, respectively, such that $(\mathbf{V})_{\alpha \beta} = (\iM-\jM) \delta_{\alpha \beta} + \jM$. Components of the impurity potential matrix $\hat{\mathbf{U}}$ is then $\hat{U}_{aa,bb} = \iM \hat{S}$ and $\hat{U}_{ab,ba} = \jM \hat{S}$. We introduce the parameter $\eta$ to control the ratio of intra- and interband scattering potentials, so that $\iM = \jM \eta$. Coupled $\Tmat$-matrix equations for $aa$ and $ba$ components of the self-energy become
\bea
\hat{\Sigma}_{aa}^{\imp} &=& n_{\imp} \hat{U}_{aa} + \hat{U}_{aa} \hat{g}_a \hat{\Sigma}_{aa}^{\imp} + \hat{U}_{ab} \hat{g}_b \hat{\Sigma}_{ba}^{\imp},
\label{eq.Sigma_aa} \\
\hat{\Sigma}_{ba}^{\imp} &=& n_{\imp} \hat{U}_{ba} + \hat{U}_{ba} \hat{g}_a \hat{\Sigma}_{aa}^{\imp} + \hat{U}_{bb} \hat{g}_b \hat{\Sigma}_{ba}^{\imp}.
\label{eq.Sigma_ba}
\eea
Renormalizations of frequencies and gaps come from $\Sigma^{\imp}_{0a} = \frac{1}{4} \mathrm{Tr}\left[\hat{\Sigma}_{aa}^{\imp} \cdot \left( \hat{\tau}_0 \otimes \hat{\sigma}_0 \right) \right]$ and $\Sigma^{\imp}_{2a} = \frac{1}{4} \mathrm{Tr}\left[\hat{\Sigma}_{aa}^{\imp} \cdot \left( \hat{\tau}_2 \otimes \hat{\sigma}_2 \right) \right]$.

Expressions for $\Sigma^{\imp}_{0\alpha}$ and $\Sigma^{\imp}_{2\alpha}$ are proportional to the effective impurity scattering rate $\Gamma_{a,b}$ and as in the case of nonmagnetic impurities contain the generalized cross-section parameter $\sigma$ that helps to control the approximation for the impurity strength ranging from Born (weak scattering, $\pi \jM N_{a,b} \ll 1$) to the unitary (strong scattering, $\pi \jM N_{a,b} \gg 1$) limits,
\bea
\Gamma_{a,b} &=& \frac{2 n_{\imp} \sigma}{\pi N_{a,b}} \to \left\{
    \begin{array}{l}
    2 \pi \jM^2 s^2 n_{\imp} N_{b,a}, \mathrm{Born}\\
    \frac{2 n_{\imp}}{\pi  N_{a,b}}, \mathrm{unitary}
    \end{array}
  \right.
\\
\sigma &=& \frac{\pi^2 \jM^2 s^2 N_a N_b}{1 + \pi^2 \jM^2 s^2 N_a N_b}
\to \left\{
    \begin{array}{l}
    0, \mathrm{Born}\\
    1, \mathrm{unitary}
    \end{array}
  \right.
\eea
We assume that spins are not polarized
and $s^2 = \la S^2 \ra = S(S+1)$.
Since $s$ enters all equations only in conjunction with $\iM$ or $\jM$, without loosing generality we set $s = 1$ assuming that $\iM$ and $\jM$ are both renormalized by $s$.

For the uniform impurity potential $\eta = 1$ in the Born limit $\sigma = 0$ we find
\bea
\tilde\omega_{an} &=& \omega_n + \ii \Sigma_{0a}(\omega_n) + \pi \jM^2 n_{\imp} \left(N_a \frac{\tilde\omega_{an}}{Q_{an}} + N_b \frac{\tilde\omega_{bn}}{Q_{bn}}\right), \nn\\ \tilde\phi_{an} &=& \Sigma_{2a}(\omega_n) - \pi  \jM^2 n_{\imp} \left(N_a \frac{\tilde\phi_{an}}{Q_{an}} + N_b \frac{\tilde\phi_{bn}}{Q_{bn}}\right). \nn
\eea
Here contribution from both $a$ and $b$ bands are mixed so we expect a suppression of $T_c$ by disorder.

In the unitary limit ($\sigma=1$) at $T \to T_c$ we have $\tilde\omega_{a n} = \omega_n + \ii \Sigma_{0a}(\omega_n) + \frac{\Gamma_a}{2} \sgn{\omega_n}$ and $\tilde\phi_{a n} = \Sigma_{2a}(\omega_n) + \frac{\Gamma_a}{2} \frac{\tilde\phi_{a n}}{\left|\tilde\omega_{a n}\right|}$ for any value of $\eta$ including the case of intraband-only impurities, $1/\eta = 0$. This form is the same as for non-magnetic impurities and thus analogously to the Anderson theorem there is no impurity contribution to the $T_c$ equation. The only exception here is the special case of uniform impurities, $\eta = 1$, when
\bea
\tilde\omega_{a n} &=& \omega_n + \ii \Sigma_{0a}(\omega_n) + \frac{n_{\imp}}{\pi  \left(N_a+N_b\right)} \sgn{\omega_n}, \nn\\
\tilde\phi_{a n} &=& \Sigma_{2a}(\omega_n) + \frac{n_{\imp}}{\pi \left(N_a+N_b\right)^2} \left(N_a \frac{\tilde\phi_{a n}}{\left|\tilde\omega_{a n}\right|} + N_b \frac{\tilde\phi_{b n}}{\left|\tilde\omega_{b n}\right|} \right). \nn
\eea
Both gaps are mixed in equation for $\tilde\phi_{a n}$, thus they tend to zero with increasing amount of disorder. That's also true away from the unitary limit and that's why there is a special case of uniform potential of the impurity scattering, $\iM = \jM$, when the strongest $T_c$ suppression occurs.

\section{Numerical results}

Following results were obtained by solving self-consistently frequency and gap equations~(\ref{eq.omega.tilde})-(\ref{eq.phi.tilde}) for both finite temperature and at $T_c$ with the impurity self-energy as in Eqs.~(\ref{eq.omega.nonmagn})-(\ref{eq.phi.nonmagn}) for the nonmagnetic disorder or from the solution of Eqs.~(\ref{eq.Sigma_aa})-(\ref{eq.Sigma_ba}) for the magnetic impurities. For definiteness we choose $N_{b}/N_{a}=2$ and coupling constants to be  $(\lambda_{aa},\lambda_{ab},\lambda_{ba},\lambda_{bb}) = (3,0.2,0.1,0.5)$ for the $s_{++}$ state and $(3,-0.2,-0.1,0.5)$ for the $s_\pm$ state with $\la \lambda \ra < 0$.

\begin{figure}[t]
\begin{center}
\includegraphics[width=.48\textwidth]{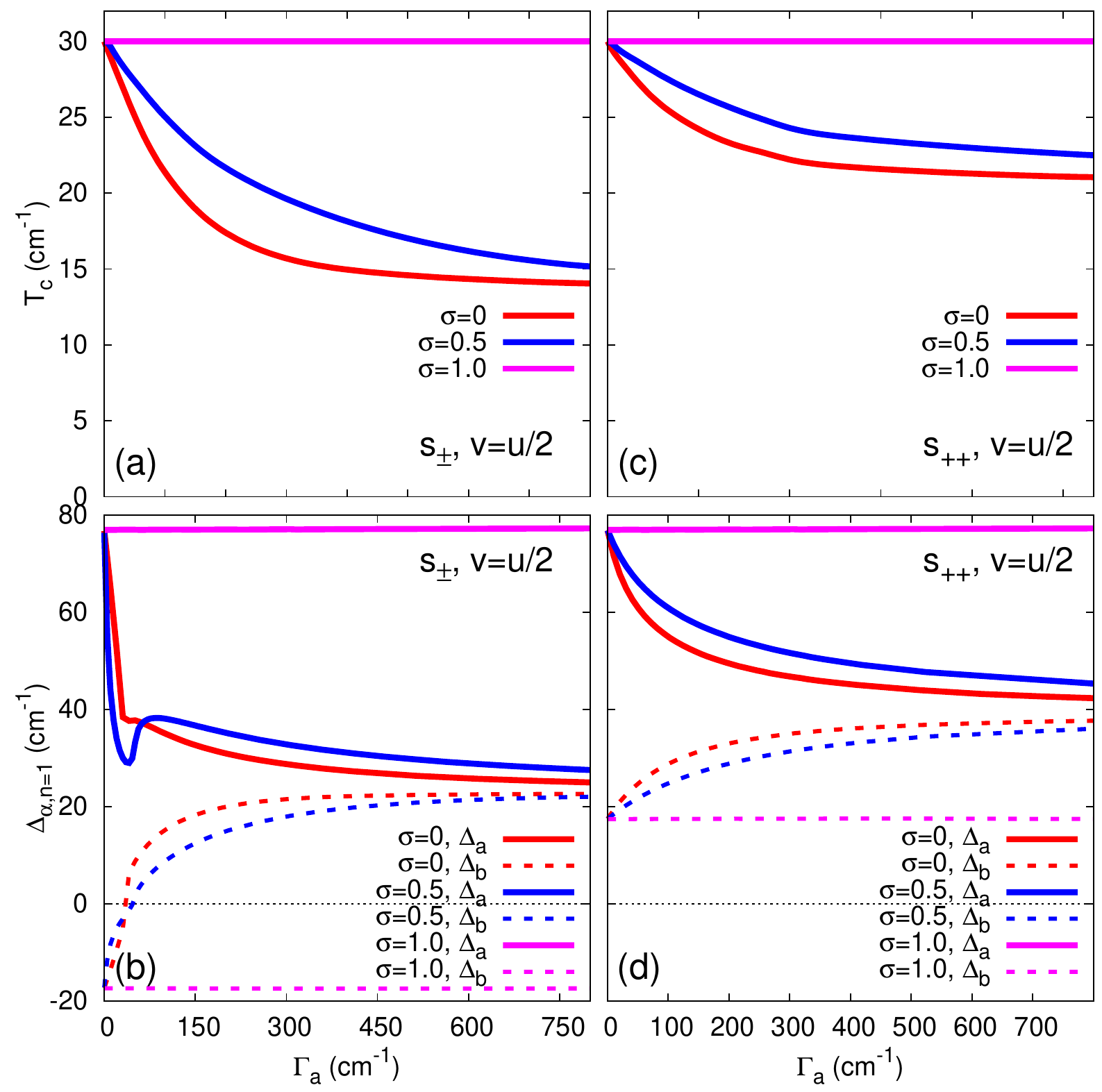}
\caption{$T_c$ (a,c) and Matsubara gap $\Delta_{\alpha n=1}$ (b,d) dependence on the nonmagnetic scattering rate $\Gamma_a$ for the $s_\pm$ (a,b) and the $s_{++}$ (c,d) superconductors with $\etaN = 1/2$.
\label{fig:spmsppTcDeltaNonmag}}
\end{center}
\end{figure}
\begin{figure}[t]
\begin{center}
\includegraphics[width=.48\textwidth]{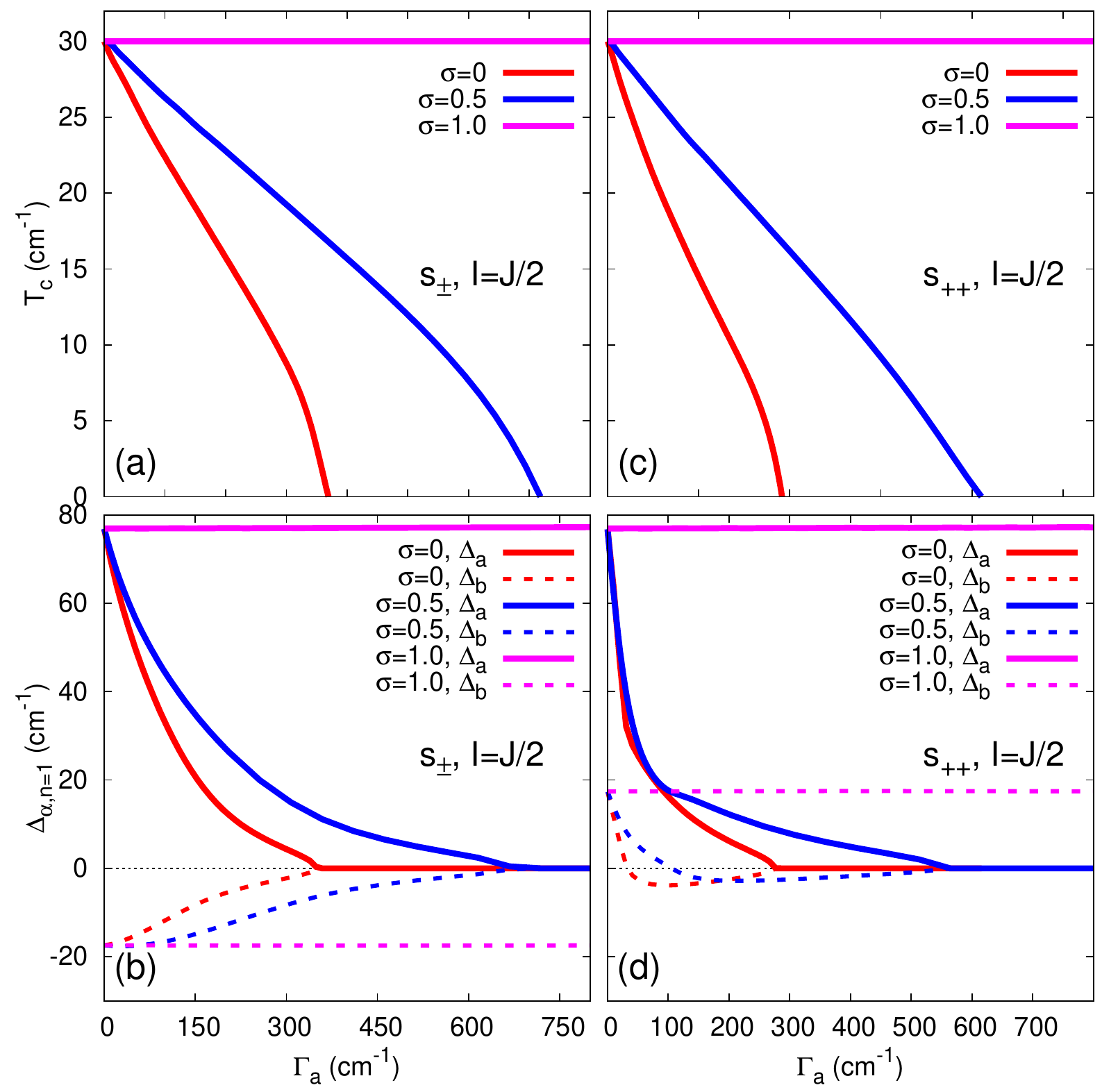}
\caption{$T_c$ (a,c) and Matsubara gap $\Delta_{\alpha n=1}$ (b,d) dependence on the magnetic scattering rate $\Gamma_a$ for the $s_\pm$ (a,b) and the $s_{++}$ (c,d) superconductors with $\eta = 1/2$.
\label{fig:spmsppTcDelta}}
\end{center}
\end{figure}

Typical results~\cite{efremov,Korshunov2014} of the dependence on the impurity scattering rate $\Gamma_a$ for the critical temperature $T_c$ and gaps $\Delta_{a,b n}$ for the first Matsubara frequency $\omega_{n=1} = 3 \pi T$ are shown in Fig.~\ref{fig:spmsppTcDeltaNonmag} (nonmagnetic) and in Fig.~\ref{fig:spmsppTcDelta} (magnetic disorder). Scattering on magnetic impurities suppress both $s_\pm$ and $s_{++}$ states due to the finite interband scattering component. The $s_{++}$ state initially transforms to the $s_\pm$ state, but then follows its fate with increasing $\Gamma_a$. The only exception is the unitary limit. On the other hand, both states survive the nonmagnetic disorder but for different reasons: the $s_{++}$ due to the Anderson theorem, while the $s_\pm$ state transforms to the $s_{++}$. Unitary limit, again, gives constant result.

\begin{figure}[t]
\begin{center}
\includegraphics[width=.48\textwidth]{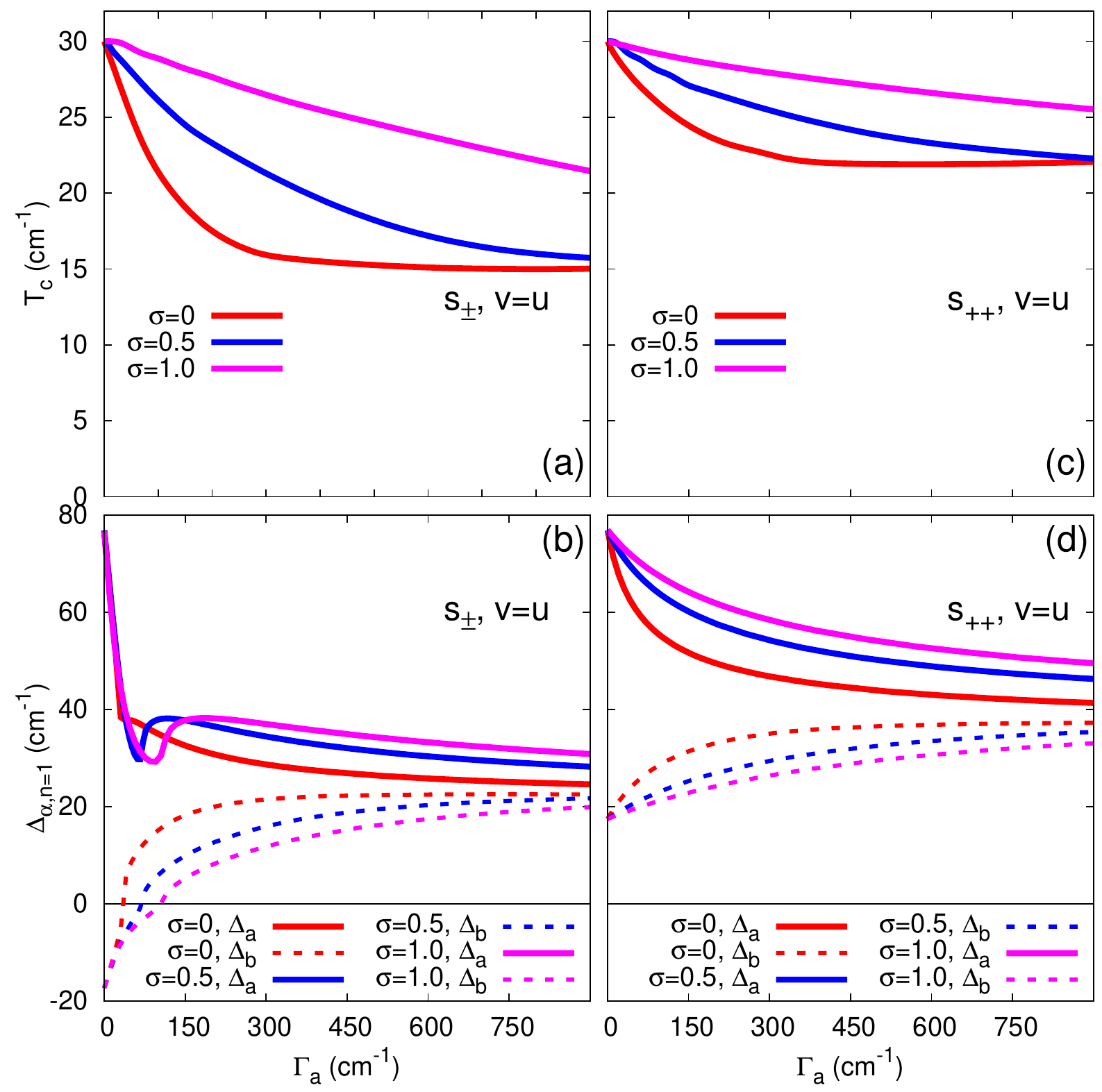}
\caption{Uniform nonmagnetic impurity potential $\etaN = 1$: $T_c$ (a,c) and Matsubara gap $\Delta_{\alpha n=1}$ (b,d) dependence on the scattering rate $\Gamma_a$ for the $s_\pm$ (a,b) and the $s_{++}$ (c,d) superconductors.
\label{fig:spmsppTcDeltaNonmag_ueqv}}
\end{center}
\end{figure}

\begin{figure}[t]
\begin{center}
\includegraphics[width=.48\textwidth]{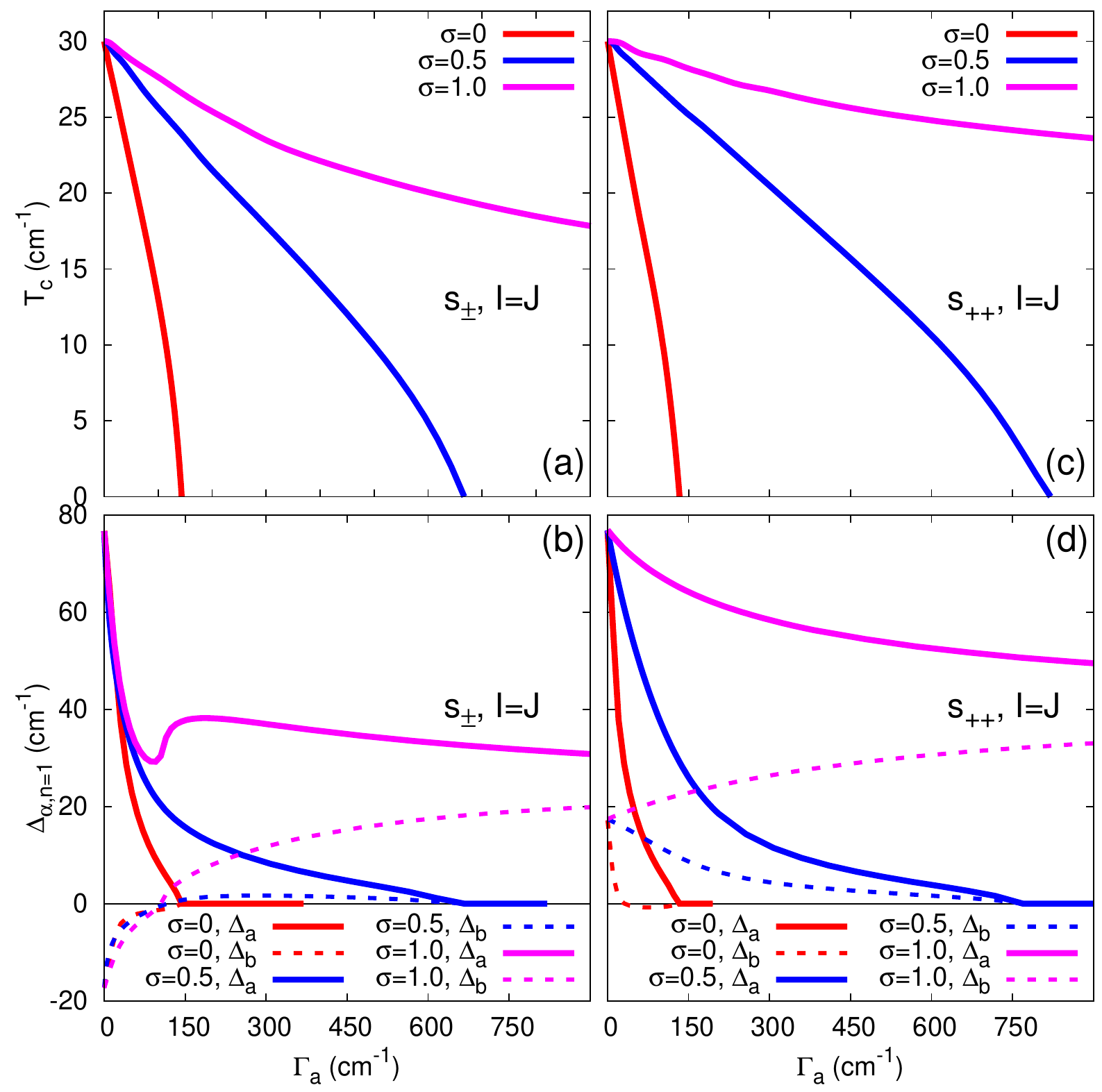}
\caption{Uniform magnetic impurity potential $\eta = 1$: $T_c$ (a,c) and Matsubara gap $\Delta_{\alpha n=1}$ (b,d) dependence on the scattering rate $\Gamma_a$ for the $s_\pm$ (a,b) and the $s_{++}$ (c,d) superconductors.
\label{fig:spmsppTcDelta_IeqJ}}
\end{center}
\end{figure}

For the uniform impurity potentials the situation, however, becomes different. Results for $T_c$ and $\Delta_{\alpha n=1}$ is shown in Fig.~\ref{fig:spmsppTcDeltaNonmag_ueqv} for the nonmagnetic disorder and in Fig.~\ref{fig:spmsppTcDelta_IeqJ} for the magnetic one. While behavior in the Born and intermediate scattering ($\sigma = 0.5$) limits are in general similar to those for $\eta \neq 1$, critical temperature and gaps in the unitary limit are not independent on disorder any more. Following the analytical results in the previous section, $T_c$ gradually decrease with increasing $\Gamma_a$. There is even a $s_\pm \to s_{++}$ transition for the magnetic impurities in the unitary limit, which is not seen for $\eta \neq 1$. On the other hand, there is no transition to the $s_\pm$ state for $\sigma = 0.5$, which appeared for $s_{++}$ state with unequal intra- and interband impurity potentials.

\section{Conclusions}

We have studied the case of uniform impurity potential, that is, the equal strength of intra- and interband scattering, $u=v$ and $\iM = \jM$ ($\eta = 1$). It appears to be qualitatively different from the other cases. This is particulary demonstrated in the unitary limit where for $\eta \neq 1$ there is an independence of gaps and $T_c$ on the values of both nonmagnetic and magnetic scattering. On the contrary, for the uniform impurity potential, there is a suppression of gaps and critical temperature due to the disorder.


\begin{acknowledgements}
We acknowledge partial support by RFBR (Grant 13-02-01395) and President Grant for Government Support of the Leading Scientific Schools of the Russian Federation (NSh-2886.2014.2).
\end{acknowledgements}

\end{document}